\documentclass[pre,showpacs,twocolumn,amsmath,amssymb]{revtex4}

\usepackage{psfrag}
\usepackage{graphicx}
\usepackage{color}
\usepackage{subfigure}
\usepackage[latin1]{inputenc}
\usepackage{amsmath}
\usepackage{amssymb}
\usepackage{relsize}

\newcommand{\mean}[1]{\left \langle #1 \right \rangle}

\begin{document}
   
\title{Optimal potentials for temperature ratchets}
  
\author{Florian Berger, Tim Schmiedl, and Udo Seifert}

\affiliation{ II. Institut f\"{u}r Theoretische Physik, Universit\"{a}t Stuttgart, 70550 Stuttgart, Germany
} 

\date{\today}

\begin{abstract}
In a spatially periodic temperature profile, directed transport of an overdamped
Brownian particle can be induced along a periodic potential. With a
load force applied to the particle, this setup can perform as a
heat engine. For a given load, the optimal potential maximizes the
current and thus the power output of the heat engine. We calculate the
optimal potential for different temperature profiles and show that in
the limit of a periodic piecewise constant temperature profile
alternating between two temperatures, the optimal potential leads to a divergent current. This divergence,
being an effect of both the overdamped limit and the infinite temperature gradient at the interface,
would be cut off in any real experiment.
\end{abstract}

\pacs{05.40.-a, 82.70.Dd}

\maketitle

\section{Introduction}
Noise induced transport occurs in a broad variety of systems both in
physics and biology \cite{reim02a}. In a microscopic system embedded in
a thermal environment, directed transport generally requires two ingredients: (i) External sources driving the system
out of equilibrium and (ii) broken spatial symmetry \cite{reim02a}. The main idea is to 
impede the thermal motion in one direction in order to obtain a net current in the
other direction \cite{astu94}. This mechanism is called the ``ratchet effect''. Applications range
from particle sorting \cite{fauc95a, rous94} to modeling molecular
motors \cite{cord92, astu94, juel97,  astu02}. 
Most studies on ratchet motors focus on a given potential landscape and
a given driving scheme.
For practical purposes, however, optimization of the driving mechanism 
with respect to  a maximal current is an important issue.

For discrete analogues of ratchet motors, where the potential landscape is characterized by only a
few parameters, such an optimization has been performed in
models for microscopic heat engines \cite {tu08} and molecular motors
\cite{schm08a}. Paradoxical Parrondo games \cite{parr00} which can be
interpreted as discrete analogues
of Brownian ratchets \cite {amen04}, have also been optimized recently
\cite{dini08}.  

For continuous motors, where optimization requires variational calculus, there exist only few results.  
The optimization of driving schemes has been studied for time-dependently driven ratchet
motors \cite{tarl98, lade08} and for a Brownian heat engine cycling between two heat reservoirs \cite{schm08}.
Potential landscapes have been optimized for the transport across
membrane channels \cite{bezr07}. Maximizing the current of flashing
ratchets by using a feedback control strategy has been proposed recently \cite{cao04, crai08}. 

In this paper,
we focus on a continuous thermal ratchet where transport along a
spatially varying time-independent potential is driven by a periodic spatial temperature 
profile  \cite{feynman, buet87, land88, blan98}.  
The recent generation of temperature gradients on small
length scales \cite{duhr06, duhr07, wein08} may render such molecular 
heat engines experimentally realizable. Cargoes driven by thermal
gradients on a subnanometer scale have already been observed \cite{barr08}.

Thermodynamic efficiencies of such ratchet heat 
engines have been calculated in Refs. \cite{jarz99a, mats00, benj08}. 
It has been argued that heat engines should generally be 
characterized by their performance 
at maximum power output \cite{curz75, vdb05}. Recent studies on ratchet
heat engines have varied the load force for a constant potential in order to
maximize the power output \cite{asfa04, gome06}. We take a complementary
approach and optimize the potential for a given load. The maximization of
power output and particle current then is equivalent.

So far, this task has been tackled numerically
only  for a one-parameter class of potentials for a given temperature \cite{asfa08}. 
To the best of our knowledge, there exists no systematic study on the
optimal potential for such a thermal ratchet. Using variational
calculus we determine the optimal 
potential which maximizes the current for a given temperature analytically up to a
numerical root search.  
 For a dichotomous temperature profile with infinitely steep
gradients as used in most previous studies \cite{asfa04, gome06, benj08}, the maximal current
diverges.  This
unphysical behaviour is an effect of the idealized assumptions of both an infinite 
temperature gradient at the interface and the overdamped dynamics. In any realistic system, 
temperature gradients will be finite which is sufficient to cut off the formal
divergence.

\section{The temperature ratchet and its current}
\begin{figure}[h!]
\centering
\includegraphics[scale=0.8]{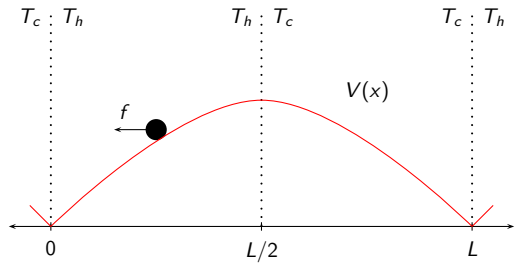}
\caption{\label{fig:model}A temperature ratchet: a particle with a load $f$ moving
  in a potential $V(x)$ in two piecewise constant temperature regions $T_h$
  and $T_c$.}
\end{figure}

We consider a Brownian particle of mass $m$ moving in a periodic
potential $V(x)$ with $V(x+L)=V(x)$ in a viscous fluid with friction
coefficient $\gamma$. A constant load $f$ is attached to the
particle. The surrounding temperature is modeled by
$T(x)$ which has the same periodicity $L$ as the
potential. A special case is a piecewise constant temperature with a hot and a
cold area, see Fig. \ref{fig:model}. For a properly chosen potential the particle moves against
the load on average. The thermal fluctuations in the hot area can
push the particle against the load over the barrier of the
potential. As soon as the particle is in the cold area the probability
of getting pushed back is smaller because of the weaker thermal
fluctuations. In this way the particle drags the load and produces
work effectively acting as a
heat engine that works between two heat baths. Such a mechanism
is not limited to piecewise constant temperatures.       

The time evolution of the position of the particle $x(t)$
is governed by the Langevin equation
\begin{equation}
m\ddot{x}= -\gamma \dot{x}- V'(x)+f+ g(x)\xi(t),
\end{equation}
where time derivatives are denoted by dots and space derivatives by primes.
The stochastic term $g(x)\xi(t)$ models the thermal noise from the
environment. Its strength
\begin{equation}
  g(x)\equiv \sqrt{{k_B T(x)}{\gamma}}
\end{equation}
depends on the temperature profile $T(x)$ resulting in
multiplicative noise. 

In the following we want to make two
assumptions: the friction dominates the inertia and the noise is
uncorrelated Gaussian.
When dealing with multiplicative noise, these two limiting procedures
do not commute \cite{sanc82, kupf04}. The order of the limiting procedures determines
the stochastic interpretation of the term $g(x)\xi(t)$. If we first assume Gaussian white
noise 
\begin{equation}
\mean{\xi(t)\xi(t')}=2\delta(t-t'),  
\end{equation}
and then go to the overdamped equation
\begin{equation}
\label{eq:Leover}
\gamma \dot{x} = - V'(x)+ f + g(x) \xi (t)  
\end{equation}
we end up with a corresponding Fokker-Planck equation
\begin{align}
  \label{eq:FP}
\partial_{t} p(x,t)
&= - \partial_{x}  \left[ \frac{1}{\gamma} \left(-V'(x)+ f \right)-
  \frac{1}{\gamma^2} \partial_{x} g^2(x) \right]p(x,t)  \\\notag
&=-\partial_{x} j(x,t)
\end{align}
 in Ito's sense where $j(x,t)$ is the current and $p(x,t)$ the probability
 distribution. 

If we first take the overdamped limit and afterwards assume Gaussian white
noise, we end
up with a Fokker-Planck equation in Stratonovich's sense. Both equations differ
in a drift term which can be absorbed in an effective potential. Thus,
the optimal current does not depend on the interpretation.
The optimal potential is merely shifted by the drift term. In the following we use
the Fokker-Planck equation in Ito's sense.

We next introduce dimensionless units. We express energies in units of
$k_B T_0$ where 
\begin{equation}
  \label{eq:T0}
  T_0 \equiv \frac{1}{L} \int_0^L T(x) \:dx. 
\end{equation}
By introducing a scaled length $\hat{x}\equiv x / L$ and a scaled
time $\hat{t} \equiv  t/t_0 $ with $t_0 \equiv L^2 \gamma / k_B
T_0$, we rewrite the Fokker-Planck equation and identify the
dimensionless quantities 
\begin{align}\notag
\hat{T}(\hat{x})&\equiv\frac{T(x)}{T_0},  &\hat{V}(\hat{x})&\equiv\frac{V(x)}{k_B
  T_0}, & \hat{f}&\equiv\frac{f L}{k_B T_0}, \\ \label{eq:diml} \hat{\jmath}(\hat{x},\hat{t})
&\equiv\frac{j(x,t) L^2 \gamma}{k_B T_0}, & \hat{p}(\hat{x},\hat{t}) &\equiv L p(x,t).   
\end{align}
For ease of notation we drop the hats in the following.
The dimensionless potential $V(x)$ and the dimensionless temperature profile $T(x)$
are periodic, $V(x+1) = V(x)$, $T(x+1) = T(x)$.
In the steady state, the current $j$ is a constant and the
Fokker-Planck equation reduces to
\begin{equation}
 j=[-V'(x)+ f- T'(x)]p(x) - T(x)\partial_x p(x).
\end{equation}
For a periodic temperature we solve this equation under the
condition of a periodic $p(x)$. Without loss
of generality, we chose $V(0)=0$ which results in
\begin{equation}
  \label{eq:prob}
p(x)=\frac{j e^{\phi(x)}}{T(x)}\left(\frac{1}{1-e^{-\phi(1)}}\int_0^1 e^{-\phi(x')}\:dx' -
  \int_0^x e^{-\phi(x')} \: dx'\right)
\end{equation}
with
\begin{equation}\label{eq:phi}
\phi(x) \equiv \int_{0}^{x}
\frac{-V'(x')+f}{T(x')} \:dx'.   
\end{equation}
Using the normalization $\int_0^1 p(x) dx =1$, we obtain the inverse
current
\begin{multline}
  \label{eq:icurrent}
j^{-1}= \frac{1}{1-e^{-\phi(1)}} \int_{0}^{1}  e^{-
    \phi(x')}\:dx'\int_0 ^1 \frac{e^{\phi(x)}}{T(x)}\:dx \\- \int_0 ^1 \left ( \int_{0}^{x}  e^{- \phi(x')}
  \:dx' \right ) \frac{e^{\phi(x)}}{T(x)}\:dx
  \end{multline}
which has been derived previously in Refs. \cite{buet87, asfa04}.

\section{Optimizing the current}
The inverse current [Eq. (\ref{eq:icurrent})] depends on the shape of the
potential $V(x)$. Instead of optimizing the current directly with respect to
the potential we introduce
\begin{equation}\label{eq:B}
B(x) \equiv \int_0^x e^{-\phi(x')}\:dx'
\end{equation}
and rewrite the inverse current in a more elegant way as a functional of $B(x)$
\begin{equation}
  \label{eq:icurrentB}
j^{-1}[B,B']= \int_0^1 \frac{\alpha - B(x)}{T(x)
  B'(x)} \: dx
\end{equation}
with
\begin{equation}
  \label{eq:alpha}
  \alpha \equiv \frac{B(1)}{1-B'(1)}.
\end{equation}
For the minimization of the functional (\ref{eq:icurrentB}), the
function space is constrained by the periodicity of the potential which
imposes a constraint on $B(x)$. With the boundary condition
$V(1)=V(0)=0$ we obtain from Eq. (\ref{eq:phi}) the non-local constraint
\begin{equation}
\label{eq:constraintphi}
\int_0^1 \phi' T\: dx = f 
\end{equation}
which by using Eq. (\ref{eq:B}) can be transformed into the isoperimetric constraint
\begin{equation}
  \label{eq:constraintB}
\int_0^1 T' \ln B'\:dx = f + T \ln B'  \Big|_0^1. 
\end{equation}
In order to minimize the inverse current [Eq. (\ref{eq:icurrentB})] under the constraint
[Eq. (\ref{eq:constraintB})] we introduce the effective Lagrangian
\begin{equation}
\mathcal{L}[B(x),B'(x),x]\equiv\frac{\alpha - B(x)}{T(x) B'(x)} + \lambda T'(x) \ln B'(x)
\end{equation}
with a Lagrange multiplier $\lambda$.

For a unique solution to the corresponding Euler-Langrange equation we have to impose two boundary
conditions. The first one, $B(0)=0$, arises naturally
from the definition of $B(x)$ in Eq. (\ref{eq:B}). One is tempted to use the condition $B'(0)=1$ as the second
one, but this is not 
appropriate. In principle, we have to allow the
derivative to jump at the boundaries because such jumps do not
contribute to the integral of the Lagrangian. 
In this way the boundary condition fixes the value $B'(0)=1$, but the value of
$\lim_{\epsilon \to 0}B'(0+|\epsilon|)$, which is the relevant boundary
condition for the solution of the Euler-Langrange equation, in fact, remains a
free parameter. This feature has been discussed in
detail for similar optimization problems \cite{schm07, gome08}.

For the optimization we
proceed in two steps. First we minimize the integral of the Lagrangian
$\int_0^1 \mathcal{L} \:dx$ and then we adjust the remaining
parameters to obtain the
maximum current.    
The corresponding Euler-Langrange equation is given by
\begin{equation}
  \label{eq:DGL}
2B'^2T+(\alpha-B)(T'B'+2TB'')+\lambda T^2 B'(T''B'-T'B'')=0 
\end{equation}
with the boundary condition $B(0)=0$. 
Changing variables
$B(x)=- \alpha \exp[I(x)]+\alpha$ leads to a second order differential equation for $I(x)$ which is integrable. The solution
\begin{equation}
  \label{eq:I}
I_{\pm}(x)\equiv   \int_0^x \frac{2 \:dx'}{-\lambda T T' \pm \sqrt{4T(c+\lambda
    T)+\lambda^2 T^2T'^2}} 
\end{equation}
still depends on the Lagrange multiplier $\lambda$ and the new
constant $c$. The solution of
 the Euler-Lagrange equation
 \begin{equation}
   \label{eq:Bsol}
 B(x)=-\alpha \exp[I_{\pm}(x)]+\alpha  
 \end{equation}
leads to the  optimal inverse current
\begin{align}
  \label{eq:minicurrent}\notag
  j^{ -1}(\lambda, c)&= -\int_0^1 \frac{dx}{T(x)I'_{-}(x)} \\
  &= \frac{1}{2} \int_0^1 \left[\sqrt{4\left(\frac{c}{T}+\lambda
      \right)+\lambda^2 T'^2} \;\right]  \: dx 
\end{align}
which depends on the two parameters $\lambda$ and $c$ but is independent of $\alpha$. Note that
in order to obtain a positive current, we chose the minus sign
of the square root in Eq. (\ref{eq:I}). 

In the next step, we optimize the inverse current [Eq. (\ref{eq:minicurrent})] 
by adjusting the free parameters $\lambda$ and $c$. These parameters are not
independent but related by the constraint
[Eq. (\ref{eq:constraintB})]
\begin{multline}
  \label{eq:constraint}
n(\lambda, c) \equiv T \ln |I_-'| \Big|_0^1+f  - \int_0^1 T' \ln |I_-'| \; dx +
\int_0^1 I_-' T \;dx
  =0.
\end{multline}
In the appendix, we show that in general
the optimization problem has a solution for $c=0$. Hence, for a
given temperature profile the remaining parameter $\lambda$ can be
determined by the constraint [Eq. (\ref{eq:constraint})].

From Eqs. (\ref{eq:phi}, \ref{eq:B}, \ref{eq:Bsol}) we derive the optimal potential
\begin{equation}
  \label{eq:optpot}
V(x)= T \ln |I_-'| \Big|_0^x  
+ fx - \int_0^x T' \ln |I_-'|\; dx' + \int_0^x I_-' T \;dx'
\end{equation}
which becomes the
basis for the following case studies.

\section{Case study I: Sinusoidal temperature profile}
\begin{figure*}
\centering
  \centering
  \subfigure[]{
    \label{fig:dt2}
 \includegraphics[scale=0.68]{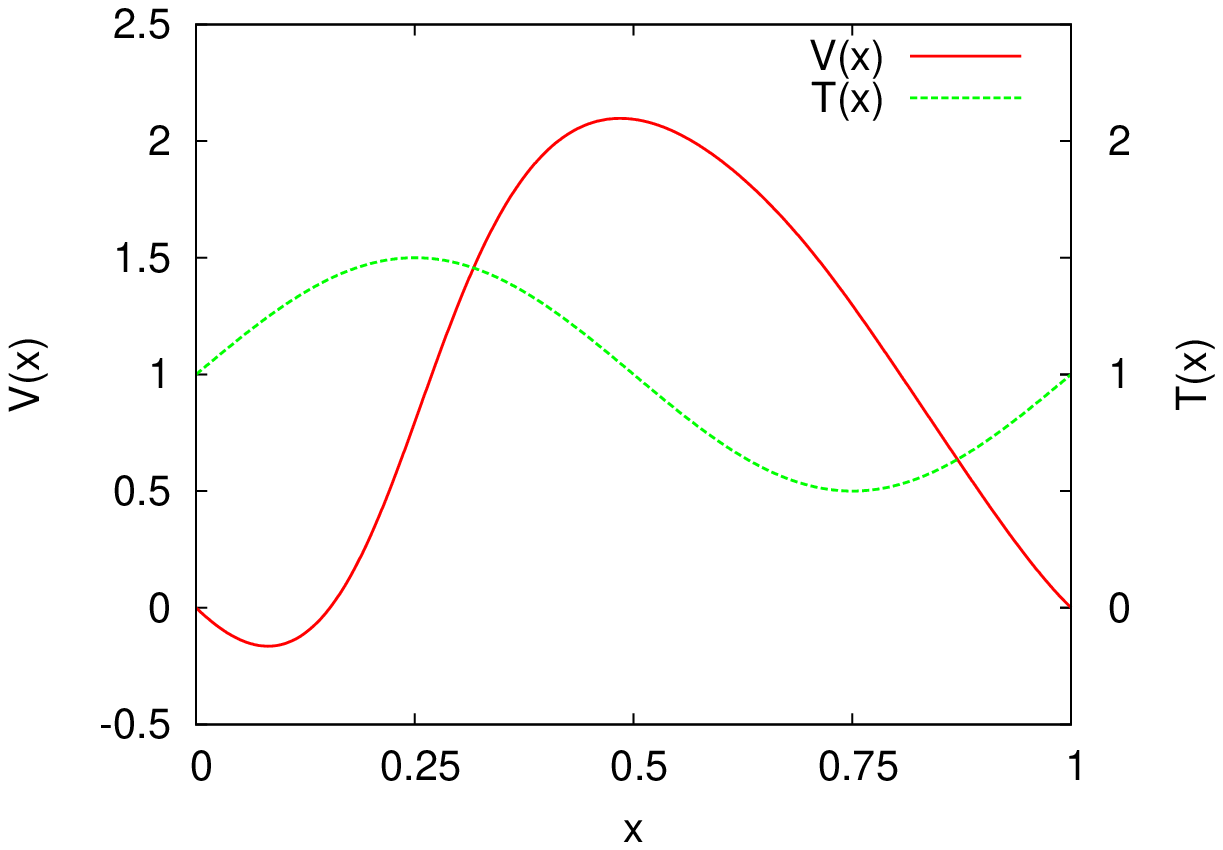}    
  }
  \subfigure[]{
    \label{fig:ampf0}
    \includegraphics[scale=0.68]{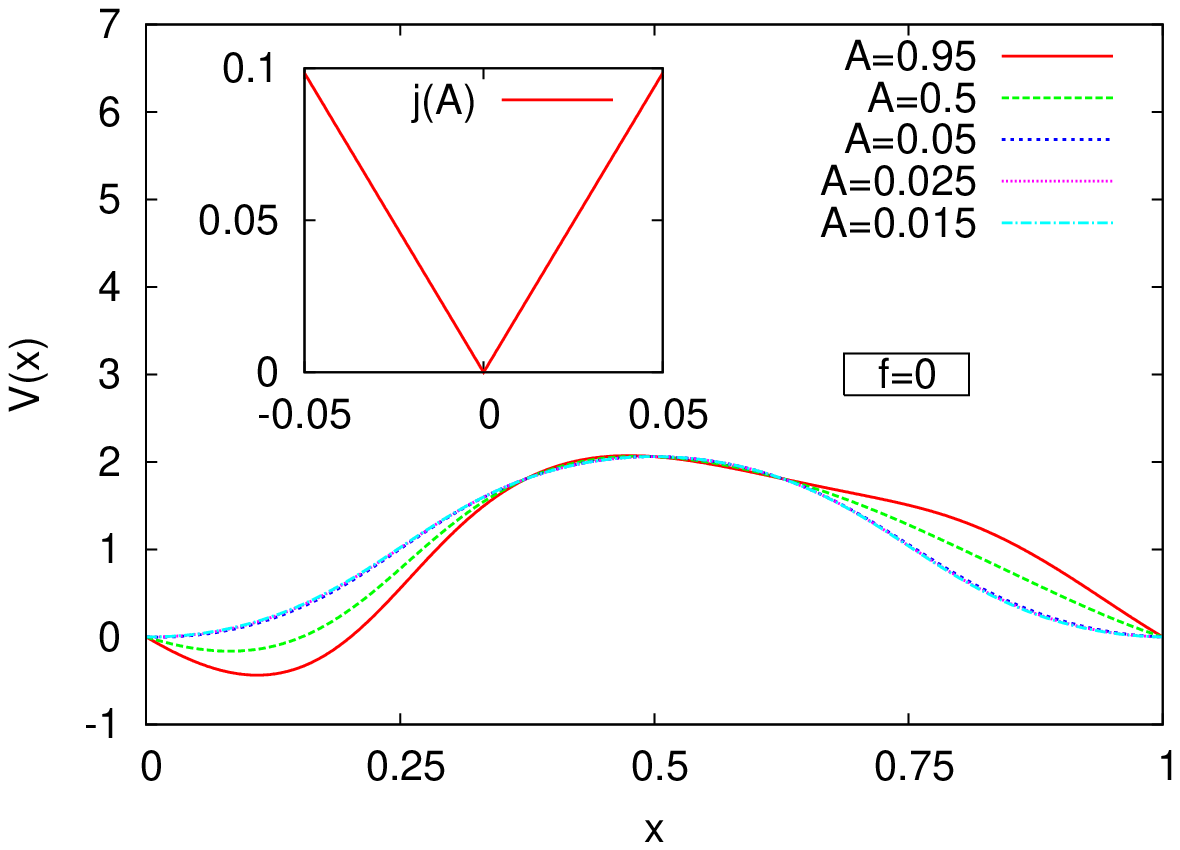}
  }
  \subfigure[]{
    \label{fig:amp}
 \includegraphics[scale=0.68]{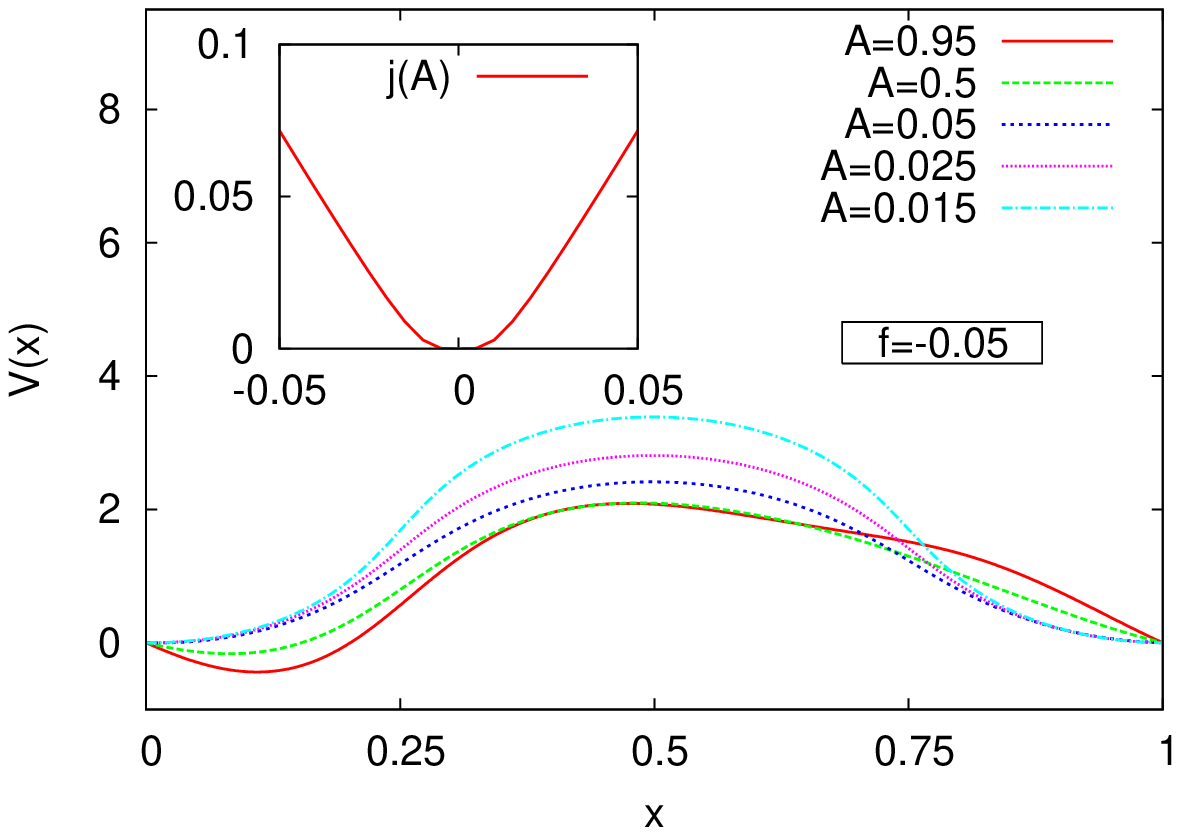}
  }
  \subfigure[]{
    \label{fig:force}
    \includegraphics[scale=0.68]{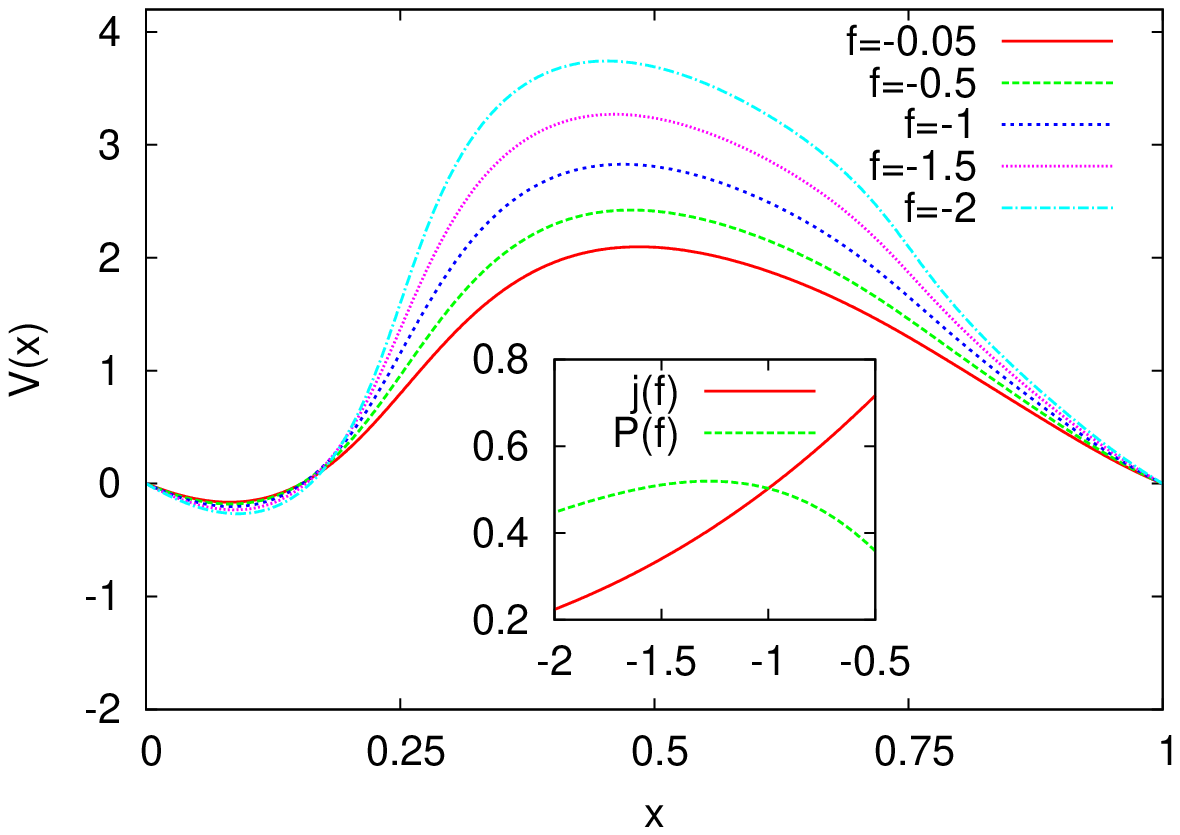}
  }
  \caption{(a) Optimal potential $V(x)$ and $T(x)=
    \frac{1}{2}\sin(2\pi x)+ 1$ for load
    $f=-0.05$. (b) Optimal potentials $V(x)$ for $T(x)= A \sin(2\pi x)+1$ with
  different amplitudes $A$ and load $f=0$. Inset: Optimal current $j$
  versus $A$. (c) Optimal potentials $V(x)$ for $T(x)= A \sin(2\pi x)+1$ with
  different amplitudes $A$ and load $f=-0.05$. Inset: Optimal current
  $j$ versus $A$. (d) Optimal potentials
  $V(x)$ for $T(x)= \frac{1}{2}\sin(2\pi x)+ 1$ and for different loads $f$. Inset:
Optimal current $j(f)$ and power $P(f)$ as a function of the force. The power exhibits a
  maximum $P_{\rm{opt}}\simeq 0.52$ at $f_{\rm{opt}}\simeq-1.27$.} 
  \label{Gemeinsames Label 1}
\end{figure*}

In this section we will discuss a sinusoidal temperature profile
\begin{equation}
  \label{eq:sintemp}
  T(x)= A \sin(2\pi x)+1
\end{equation}
with the amplitude $0 < A < 1$. The external
force $f$ is a second parameter. In the following we will study how the
optimal potential and the current depend on these two parameters.

The height of the potential is the essential blocking mechanism in
the ratchet. The thermal ``kicks'' from the environment move the
particle over this barrier. We expect the largest slope of the
optimal potential roughly at the hottest point because there the fluctuations are strong enough
to push the particle against a large force. In the colder regions, the potential
should decrease. The optimal potential as determined numerically
using Eq. (\ref{eq:optpot}) indeed fulfills these
expectations, see Fig. \ref{fig:dt2}. 
 
\subsection{Optimal potential for different amplitudes of the
  temperature profile}

The optimal potential for different amplitudes $A$ with zero
external force $f=0$ is shown in Fig. \ref{fig:ampf0}.
For amplitudes $A\rightarrow 1$, the temperature in the colder area $x>0.5$ goes
to zero and the optimal potential becomes strongly asymmetric. In
this low temperature area the thermal fluctuations are so weak that
even a gently declining potential is sufficient to push the particle
in one direction.  
The optimal current is proportional to the absolute value of the amplitude, see
Fig. \ref{fig:ampf0}. This general scaling behavior is not limited to a
sinusoidal temperature profile which can be understood as follows. With the periodicity $T'(0)=T'(1)$,
the first term of the constraint [Eq. (\ref{eq:constraint})] vanishes. With $c=0$ and $f=0$,
Eq. (\ref{eq:constraint}) and Eq. (\ref{eq:minicurrent}) become  
\begin{multline}
  \label{eq:constraintamp}
 n(\lambda) =  \int_0^1 T' \ln\left|\lambda T' +
   \sqrt{4\lambda+\lambda^2 T'^2}\right| \\- \frac{1}{2 \lambda} \sqrt{4\lambda
   +\lambda^2 T'^2}\:dx = 0
  \end{multline}
and
\begin{equation}
  \label{eq:currentamp}
  j^{ -1}(\lambda)= \frac{1}{2} \int_0^1 \sqrt{4\lambda +\lambda^2 T'^2}  \: dx.\end{equation}
The derivative of the temperature scales like $T'(x) \propto A$. Choosing $\lambda
\propto A^{-2}$, the constraint
[Eq. (\ref{eq:constraintamp})] is independent of $A$. The current
[Eq. (\ref{eq:currentamp})] then is a linear function of $A$. 

Now we consider the system with a finite external force, $f=-0.05$.
For temperature amplitudes $A\rightarrow 1$, effectively corresponding
to a large temperature difference, the external force can be neglected
and the optimal potential looks like in the case without an external
force, see Fig. \ref{fig:amp}.
For small $A$,
the ratchet effect induced by the temperature difference is not strong
enough against the external force. In this regime, the optimal
potential has to prevent the particle from being dragged in the
direction of the external force. This is achieved by blocking the
particle with a larger barrier, see Fig. \ref{fig:amp}. 

\subsection{Optimal potential for different external forces}
For stronger external forces, the potential has to compensate this
dragging mechanism with a larger barrier in order to obtain a current
in the direction opposite to the force. Thus, we expect larger potentials and smaller currents for stronger
external forces which is confirmed by our calculations, see Fig. \ref{fig:force}.

We next calculate the dimensionless power output of the heat engine which is given by
\begin{equation}
  \label{eq:power}
  P\equiv - f j.
\end{equation}
The power output as a function of the force $f$ is shown in the inset in Fig. \ref{fig:force}.
It exhibits a maximum at intermediate forces where the heat engine thus works in a maximum 
power regime.

\begin{figure*}
\centering
  \centering
  \subfigure[]{
    \label{fig:sinap}
\includegraphics[scale=0.68]{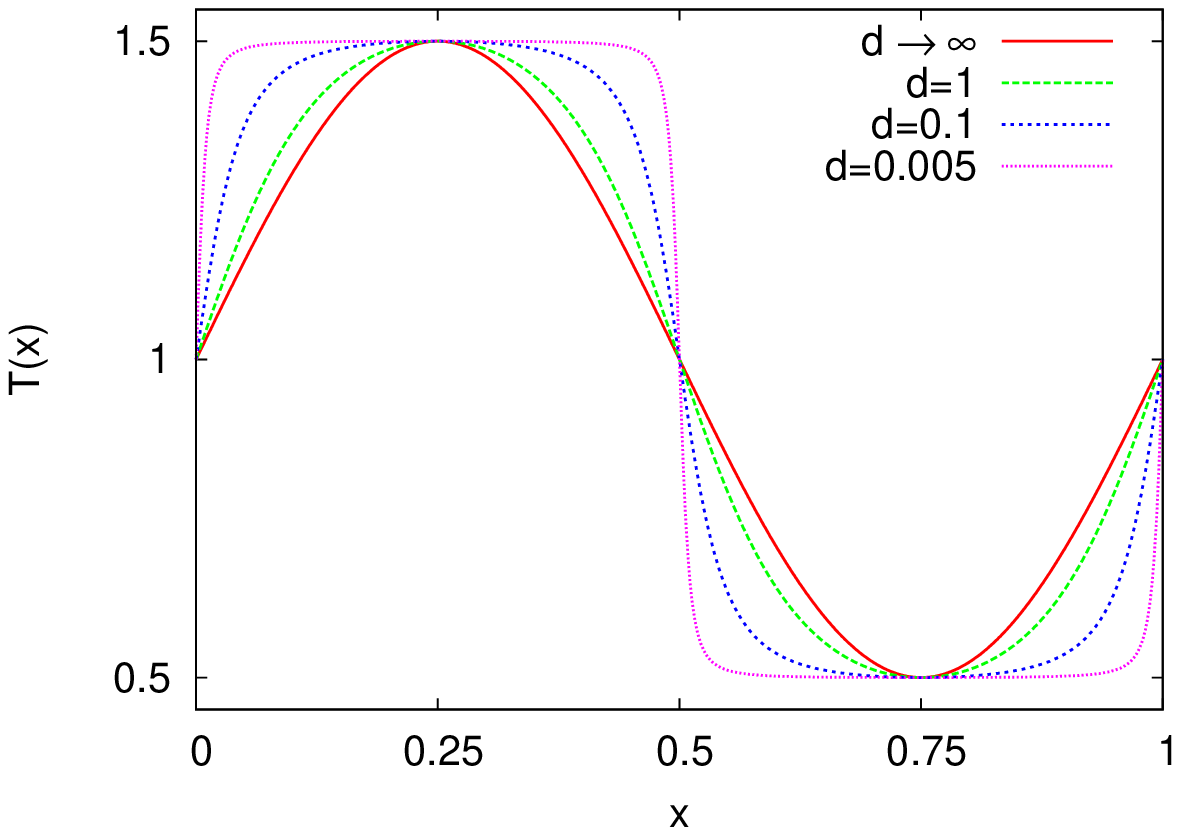}
  }
  \subfigure[]{
    \label{fig:pss}
\includegraphics[scale=0.68]{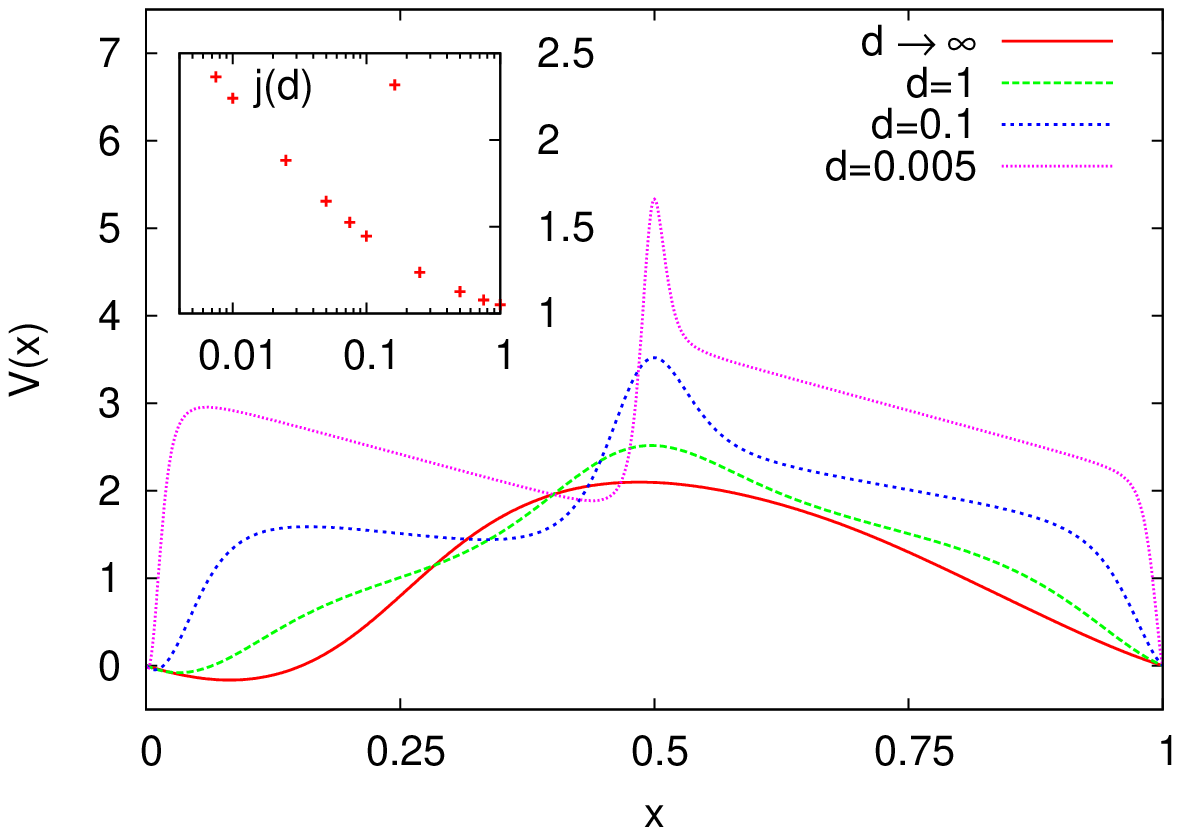}
  }
  \subfigure[]{
    \label{fig:numpot}
\includegraphics[scale=0.68]{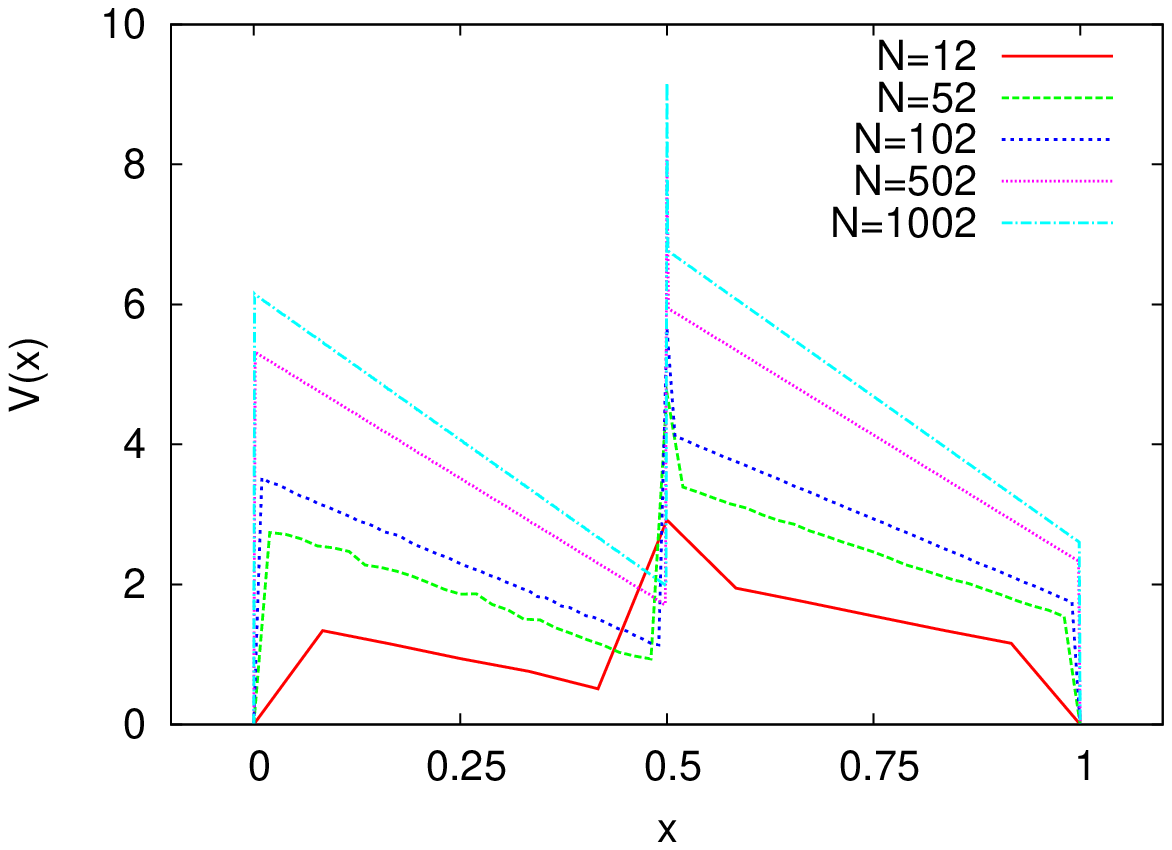}
  }
  \subfigure[]{
    \label{fig:pert}
    \includegraphics[scale=0.68]{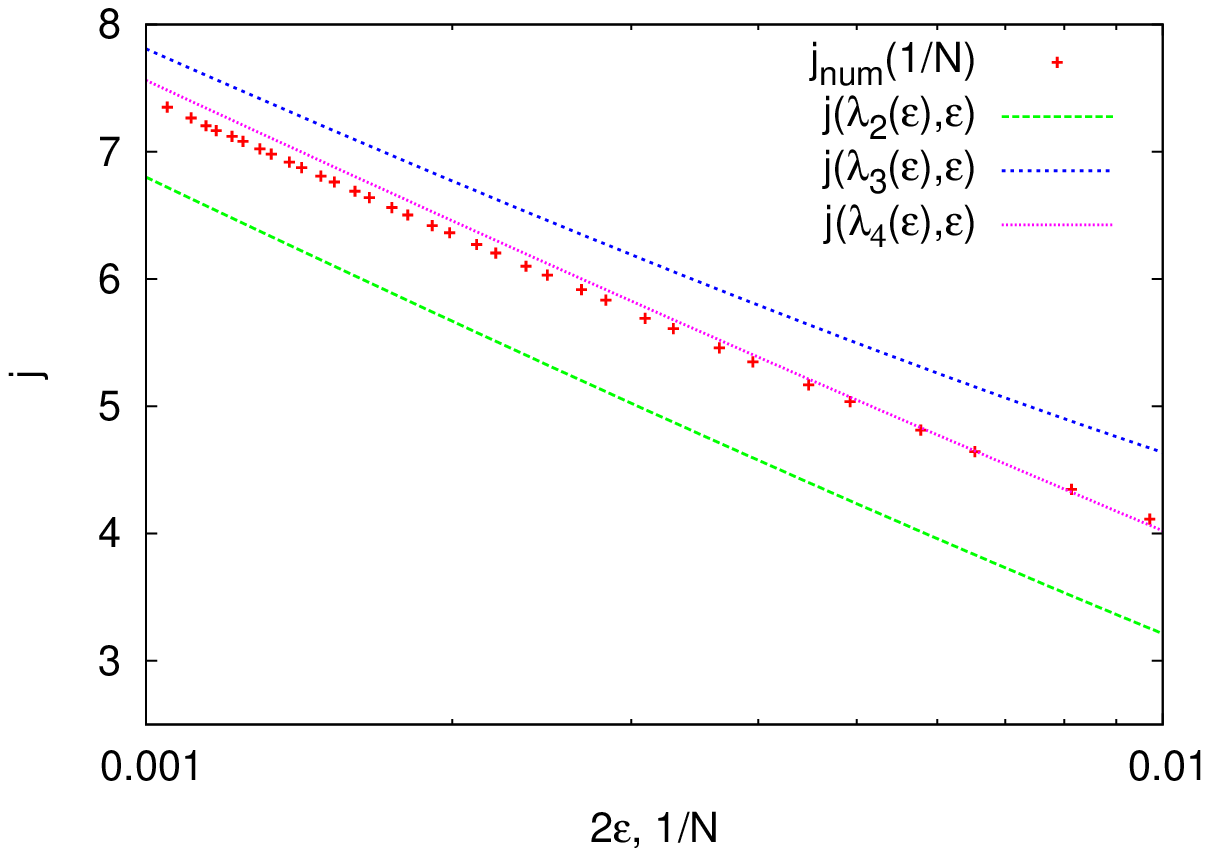}
  }

  \caption{(a) Continuous
    approximation $T(x)$ [Eq. (\ref{eq:contemp})] to a piecewise constant temperature for different parameters
    $d$. In the limit $d \to \infty$, $T(x)$ approaches the sinusoidal profile $T(x)= \frac{1}{2}\sin(2\pi x)+ 1$. (b) Optimal
    potentials $V(x)$ for $T(x)$ from Fig. \ref{fig:sinap} with
    external force $f=-0.05$. Inset: current $j$ versus parameter $d$. (c) Optimal potentials obtained
  from the numerical minimization on a discrete lattice for a
  piecewise constant temperature profile $T(x)$ [ Eq. (\ref{eq:jumpT})]
  with $\Delta T=1/2$ and with an external force $f=-0.05$ for different
  discretization $N$. (d) For an external force $f=-0.05$, optimal current $j_{\rm{num}}$ as obtained
  from the
  numerics with discretization $2 \epsilon = 1/N$ compared to the perturbative calculation developed in the appendix.}
  \label{Gemeinsames Label 2}
\end{figure*}

\section{Case study II: Piecewise constant temperature}
We now consider a piecewise constant temperature
\begin{equation}
  \label{eq:jumpT}
  T(x)= 1+\Delta T -2\Delta T\Theta(x-1/2)
\end{equation}
 with a hot and a cold area with temperatures $T_h \equiv 1+\Delta T$
 and $T_c \equiv 1-\Delta T$, respectively, where $0 <
 \Delta T < 1$. We then face discontinuities in the Fokker-Planck equation. In this section,
we first analyze a continuous approximation to the piecewise constant
temperature. The optimal potential then has an complex shape with peaks. We compare 
these results with a numerical solution for the optimal potential. 

The continuous temperature profile
\begin{equation}
  \label{eq:contemp}
T(x)=\frac{\sqrt{1+d}\sin(2 \pi x)}{2 \sqrt{\sin^2(2 \pi x)+d}} +1
\end{equation}
interpolates between the extreme values $d \to 0$ which corresponds to the
piecewise constant temperature [Eq. (\ref{eq:jumpT})] with
$\Delta T=1/2$ and $d \to \infty$ where the profile
becomes sinusoidal [Eq. \ref{eq:sintemp}] with $A=1/2$, see Fig. \ref{fig:sinap}.

The optimal potential for different values of the parameter $d$ is
shown in Fig. \ref{fig:pss}.
For $d \ll 1$, two peaks  
emerge in the optimal potential at the positions of the temperature discontinuities. In between
these two peaks, the potential decreases linearly. The
current diverges for $d \to 0$, see inset of Fig. \ref{fig:pss}.

\subsection{Numerical solution for a piecewise constant temperature}   
We next investigate this complex shape of the potential and the divergent current in more
detail by solving the problem numerically on a discrete lattice.
The goal is to minimize the inverse current [Eq. (\ref{eq:icurrent})] for
the piecewise constant temperature [Eq. (\ref{eq:jumpT})]. The periodic boundary condition
[Eq. (\ref{eq:constraintphi})] for the potential transforms to the condition
\begin{equation}
  \label{eq:conditionphi}
  \phi(1) =\phi(1/2)\left(\frac{2 \Delta T}{\Delta T -1}\right)
  +\frac{f}{1-\Delta T}  
\end{equation}
considering that $\phi(0)=0$ by definition [Eq. (\ref{eq:phi})]. 
For a numerical solution we discretize
$\phi(x)\rightarrow\phi(x_i) \equiv \phi_i$ with $i=0,\dots ,N$ and search for a global minimum
of $j^{-1}(\phi_i)$ in this $N+1$-dimensional space with
a simplex algorithm proposed by Nelder and Mead \cite{neld65}. The
boundary values $\phi_0$ and $\phi_{N}$ are given according to
Eqs. (\ref{eq:phi}, \ref{eq:conditionphi}) by
\begin{align}
  \phi_0&=0,\\
  \phi_{N}&=\phi_{N/2}\left(\frac{2 \Delta T}{\Delta T -1}\right)
  +\frac{f}{1-\Delta T}  
\end{align}
and $\phi_1,\dots ,\phi_{N-1}$ are varied to yield a maximum current. Using
Eq. (\ref{eq:phi}), we then calculate the potential $V(x)$ from the optimal $\phi(x)$.

The numerical solution for the optimal potential depends on the
discretization which determines how large the gradient
of the temperature and the potential can be, see
Fig. \ref{fig:numpot}. For finer discretization, the optimal potential
shows larger gradients. As a consequence, we find that the current $j_{\rm{num}}(N)$ diverges with increasing
discretization $N \to \infty$, see Fig. \ref{fig:pert}. This is consistent with the
 developing divergence visible in Fig. \ref{fig:pss}.  

\subsection{Origin of the divergent current}
 \begin{figure}[h!]
 \centering
\includegraphics[scale=0.8]{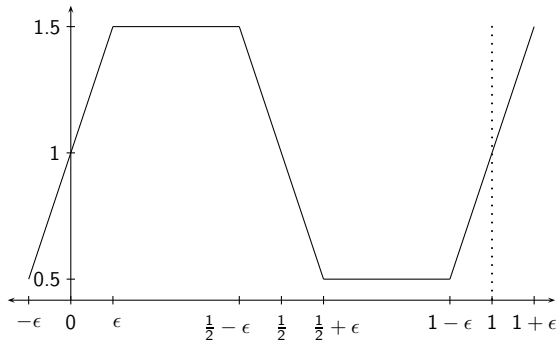}
 \caption{Temperature profile on the lattice with gradients depending
   on $\epsilon$.}
 \label{fig:templat}
 \end{figure}
Due to the lattice discretization, the temperature
profile can be considered to be linear with gradients with absolute
value $1/(2\epsilon)$, see Fig. \ref{fig:templat}.
We calculate the optimal inverse current [Eq. (\ref{eq:minicurrent})] for
such a temperature profile 
\begin{equation}
  \label{eq:pminicurrent}
  j^{-1}(\lambda,\epsilon)= \sqrt{\lambda^2 + 16 \lambda \epsilon^2} + \sqrt{\lambda}(1-4\epsilon).
\end{equation}
The Lagrange multiplier $\lambda$ and the discretization parameter $\epsilon$ are related by the corresponding constraint
[Eq. (\ref{eq:constraint})] 
\begin{equation}
  \label{eq:pconstraint}
  \ln\left|1+\frac{\lambda}{8\epsilon^2}+\frac{1}{4
      \epsilon}\sqrt{\frac{\lambda^2}{4 \epsilon^2}+4
      \lambda}\right|-\sqrt{1+\frac{16
        \epsilon^2}{\lambda}}+\frac{4\epsilon-1}{\sqrt{\lambda }}+f=0.
\end{equation}
In the appendix, we show with a
perturbation method that for $\epsilon \rightarrow 0$, the constraint
[Eq. (\ref{eq:pconstraint})] can be fulfilled if $\lambda \rightarrow 0$.
For an approximated dependence $\lambda(\epsilon)$, we use an
iterative method and obtain a sequence of functions 
\begin{equation}
  \lambda_{n+1}(\epsilon) =\frac{1}{(\ln|\lambda_n(\epsilon)|-2\ln|2 \epsilon|-1+f)^2}
\end{equation}
with $\lambda_0=1$. For the first members we calculate the current [Eq. (\ref{eq:pminicurrent})] as a
function of $\epsilon$. These results are
compared to the numerical current, where the discretization $1/N$
corresponds to $2 \epsilon$, see Fig. \ref{fig:pert}.   
The divergent behavior of the current obtained from perturbation theory is in good agreement with the divergence from the numerics. 
The discrepancy is
on one hand due to the numerical integration on the lattice in
Eq. (\ref{eq:icurrent}) and on the other hand due to the
approximations implicit in the perturbation method. 

\begin{figure*} 
\centering
  \centering
  \subfigure[]{
    \label{fig:peakdis}
\includegraphics[scale=0.68]{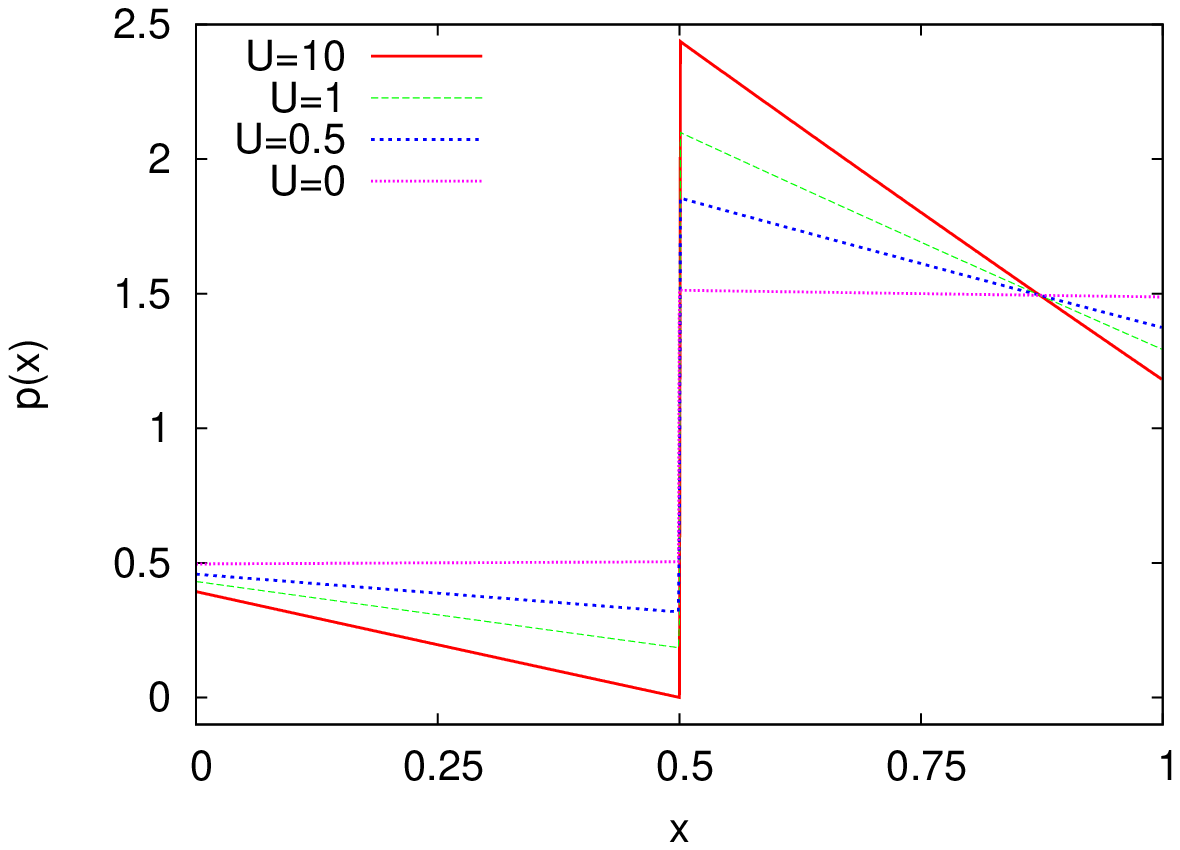}
  }
  \subfigure[]{
    \label{fig:sawpeak}
\includegraphics[scale=0.68]{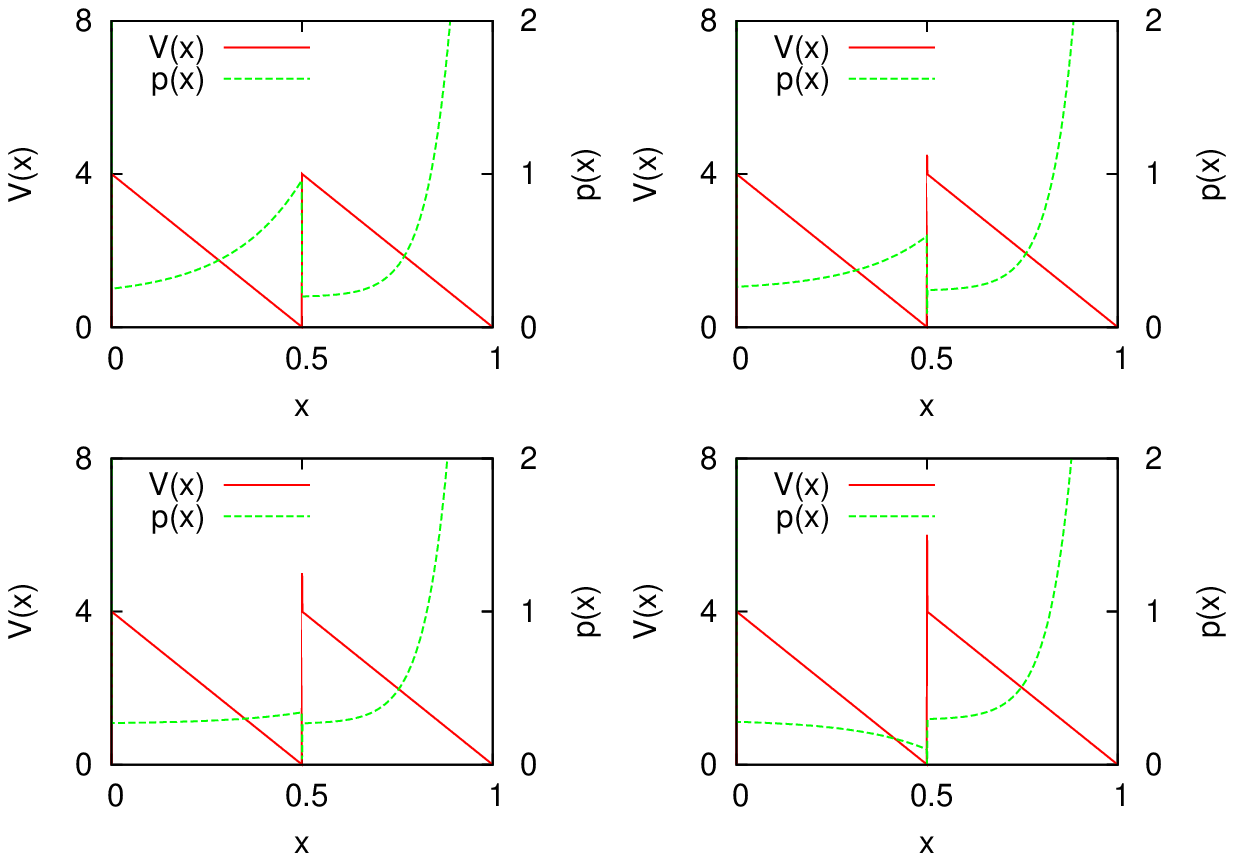}
  }
  \caption{(a) Probability distributions for a rising peak with height
    $U$ in a piecewise constant temperature $T(x)$ [Eq. (\ref{eq:jumpT})] with
    $\Delta T=1/2$ and load $f=-0.05$. (b) Sawtooth potential
    with superimposed peak and its corresponding probability
  distribution in a piecewise constant temperature $T(x)$
  [Eq. (\ref{eq:jumpT})] with
    $\Delta T=1/2$ and load $f=-0.05$ for different peak heights.}
  \label{Gemeinsames Label 3}
\end{figure*}

More insight into the origin of the divergent current can be gained
from analyzing the probability distribution [Eq. (\ref{eq:prob})]
which can be calculated from the Fokker-Planck equation for a model potential.
Guided by the optimal potential obtained from the numerics, see Fig. \ref{fig:numpot}, we consider
a sawtooth potential with a superimposed finite peak at the
temperature discontinuity at $x=0.5$. Note that we use this potential only as a case study aiming at
a deeper understanding of the divergence of the current. The full optimal potential has
a second peak at $x=0$ and both the height of the peaks and the linear slopes between
the peaks diverge.
\begin{figure}[h!]
\centering
\includegraphics[scale=0.8]{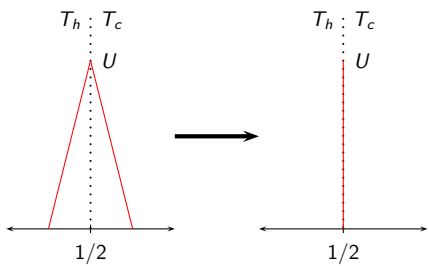}
\caption{Peak as a limiting process. Each side is in one temperature region.}
\label{fig:peaklimit}
\end{figure}
A finite peak potential is not a $\delta$-function, but can be
considered as a result of a limiting process from a triangular shape, see
Fig. \ref{fig:peaklimit}. It is
important that the peak arises at the discontinuity of the temperature
profile, namely that the rising side of the peak is in one temperature area and
the other side is in the other temperature area. Otherwise the peak
would not contribute to the
integrals in the current. Since the peak has infinitesimally
  small width, the forward rate is not decreased
  by a higher peak potential, contrary to a rising barrier with
  finite width.  
A finite peak
in a constant temperature area would not contribute to integrals and
thus would have no effect on mean first passage times. In contrast, for a dichotomous temperature
profile and a finite peak at the interface, mean first passage times involving a crossing of the peak are affected by its height.
 The backward mean first passage time increases exponentially with the peak height. In contrast and somewhat counter-intuitively, the forward mean first passage time decreases with increasing peak height due to the strong suppression of peak recrossings from the cold side. 

A finite peak with height $U$ superimposed on a flat potential causes a depletion of the particles on the high
temperature side and an accumulation on the other side. With increasing
peak height $U$, this behavior saturates, see Fig. \ref{fig:peakdis}.
In a sawtooth potential, the particles accumulate in
front of the barriers, see Fig. \ref{fig:sawpeak}.
If we combine both effects by superimposing a finite peak on a
sawtooth potential, the depletion
can compensate the accumulation, see Fig. \ref{fig:sawpeak}. 

The compensation of the accumulation is the main reason for the divergent current. For a sawtooth potential with
large amplitude, the
particles would accumulate in front of the barriers. The peaks
counteract this effect and allow the particles to overcome the barriers. In between
the peaks, the particles (on average) follow the steep
potential with large mean velocity. For
finer discretization, the peaks get larger and the potential in between
steeper corresponding to a stronger force which increases the mean velocity and thus the current.

This behavior is only valid under the assumption of an
  overdamped dynamics where the particle instantaneously adjusts its velocity according to the Langevin equation (\ref{eq:Leover}). In the underdamped case, the particle needs a
  certain distance to reach the velocity corresponding to the local force. The overdamped
  limit is a good approximation as long as the relaxation time of the momentum
  $\tau_R \equiv m / \gamma$ is small compared to the (mean) time $\tau_c$ it takes for
  the particle to cross a region with basically constant force. Considering
  the potential above with discretization $\epsilon$, both the peaks and the linear slopes
  between the peaks should not become too large for the overdamped limit to be appropriate.
  The peaks constitute the largest slope in the optimal potential and therefore are crucial
  for the appropriateness of the overdamped description.
  Considering a peak with height $U$ and width $2 \epsilon L$, 
  the relaxation time $\tau_R$ should be small compared to $\tau_c =
  \epsilon L / v^s$ where $v^s = |V' / \gamma| = U / (\epsilon L \gamma)$ is the (mean) stationary
  velocity. Therefore, the overdamped limit is approriate for
  \begin{equation}
    \frac m \gamma \ll \frac {L^2 \epsilon^2 \gamma} {U} ~~ {\rm or} ~~ \frac m {\gamma^2 L^2} \ll \frac {\epsilon^2} {U}.
    \label{overd}
  \end{equation}
  For a given value of $m/(\gamma^2 L^2)$, the discretization
  $\epsilon$ thus cannot become too small for the overdamped limit to
  be still valid.
  Since the divergence of the current occurs with decreasing values of $\epsilon$, the current 
  presumably does not diverge in any 
  realistic system, where the underlying dynamics is underdamped. This question, however, is hard to decide conclusively, since
  the current for underdamped dynamics can only be determined numerically for a given potential.
  The optimization of the potential (with an infinite number of
  degrees of freedom) is 
  a computationally difficult task to be reserved to future work.
  For a colloidal particle of radius $R \simeq 1 \rm{\mu m}$ in a
  temperature profile with $T_0 \simeq 300 K$ and periodicity $L
  \simeq 50 \rm{\mu m}$, we get
  \begin{equation}
    \frac {m} {\gamma^2 L^2} \simeq 2 \cdot 10^{-11} \frac 1 {k_B T_0}
  \end{equation}
  where we have used Stokes friction $\gamma = 6 \pi \eta R$ and the
  mass $m = 4 \pi \rho R^3 / 3$ with viscosity $\eta \simeq 10^{-3}
  \rm{\,Pa\; s}$ and density $\rho \simeq  10^3 \rm{\, kg /
    m^3}$. Even for the finest discretization $\epsilon \simeq 5 \cdot 10^{-4}$
  used in our study, we have $\epsilon^2 / U \simeq 3 \cdot 10^{-8} / {k_B T_0}$
  and condition (\ref{overd}) is fulfilled. The overdamped description thus
  is valid for the optimal potential for any realistic (finite)
  temperature gradient. The (large) currents shown in
  Fig. \ref{fig:pert} thus are in principle observable in experiments,
  provided the necessary temperature jump can be generated on the
  scale of $2 \epsilon \cdot  L \simeq 10^{-3} \cdot 50  \rm{\mu m}
  \simeq 50  \rm{nm} $. A genuine divergence of the current, on the
  other hand,
  may be prohibited by the onset of inertia effects.

\section{Conclusions}
Using variational calculus, we have developed a method to calculate the 
optimal potential which maximizes the current in ratchet heat engines for a continuous 
temperature profile. 
In the load free case, we have shown that the maximum current 
depends linearly on the amplitude of the temperature profile. 

In the case of a piecewise constant temperature, the current diverges for the
optimal potential consisting of steep linear parts and peaks at the
boundaries which induce a long range effect in the probability
distribution. However, the underlying assumption of an overdamped
dynamic is limited by the slopes of the optimal potential presumably resulting in
a bounded current for a particle with finite mass. In addition, a piecewise constant 
temperature is an artificial model rather than physically feasible. For any future nano- or 
microfluidic realization, the temperature gradients will be finite and thus the current.

In principle, external potentials for a Brownian particle 
can be realized by laser traps. However, it might be very
difficult to model the optimal potential in every detail.
In particular for the dichotomous temperature profile, it 
is clear that finite peaks must
be approximated by a barrier with a finite width in any experiment. In
order to estimate the observable velocity, we consider a
colloidal particle with radius $R$ trapped in the optimal potential
for a sinusoidal temperature profile. In recent experiments,
temperature gradients $\Delta T / L \simeq 10^{5}\; \rm{K/m}$ have
been generated \cite{wein08}. In the load free case, the
optimal dimensionless current is roughly $\hat{\jmath} \simeq 2 A$, see
Fig. \ref{fig:ampf0}, where $A$ is
the scaled temperature amplitude $\Delta T / T_0$. With
Eq. (\ref{eq:diml}) we get a rough estimate for the stationary
velocity
\begin{equation}
  \label{eq:statv}
  v\simeq \frac{k_B \Delta T}{3 \pi \eta R L}\simeq\frac{100}{R/\rm{nm}} \frac{\rm{nm}}{\rm{s}},
\end{equation}
under the assumption of Stokes friction with viscosity $\eta \simeq 10^{-3}
  \rm{\,Pa\; s}$. Although such a transport effect is small at a micrometer scale, future realisations at a nanometer scale
will yield observable velocities.

Our study can be extended in several directions. In principle, it would be interesting to calculate the efficiency at maximum power for our 
model and compare it to previous results where only the load force was optimized. 
However, it is difficult to define efficiencies for continuous
temperature profiles. Moreover, Hondou and Sekimoto pointed out in Ref.
\cite{hond00} that heat transfer cannot be treated appropriately within the overdamped Langevin
equation. 
For a similar model of a ratchet heat engine, a heuristic argument was used to propose 
a potential which leads to a large Peclet number \cite{lind02}. It would be interesting to see whether our 
approach can be used to calculate the optimal potential with respect to a maximal Peclet number.

While ratchet heat engines have not been realized experimentally yet,
the recent successful generation of large temperature gradients
\cite{duhr06, duhr07, wein08} may facilitate the construction
of microscopic heat engines. By using our results, such nanomachines may 
then be tuned to produce maximum power.

\begin{acknowledgments}
We would like to thank A. Gomez-Marin and R. Finken for inspiring discussions.
\end{acknowledgments}

\section{Appendix}
\subsection{Optimization of the current with respect to the remaining
  free parameters $\lambda$ and $c$}
With $I_-(x)$ from Eq. (\ref{eq:I}), the inverse current
[Eq. (\ref{eq:minicurrent})] reads
\begin{equation}
  j^{ -1}(\lambda, c)= \frac{1}{2} \int_0^1 \left[\sqrt{4\left(\frac{c}{T}+\lambda
      \right)+\lambda^2 T'^2} \;\right]  \; dx 
\label{eq:curentmini}
\end{equation}
with the constraint [Eq. (\ref{eq:constraint})] for $\lambda$ and $c$
 \begin{multline}
 n(\lambda,c) =  \int_0^1 T' \ln\left|\lambda T' +
   \sqrt{4\left(\frac{c}{T}+\lambda\right)+\lambda^2 T'^2}\right|  \:
 dx \\+ \int_0^1 \frac{-2 \: dx}{\lambda T' +
     \sqrt{4(\frac{c}{T}+\lambda )+\lambda^2 T'^2}} +f =0.
\label{eq:constraintmini}
 \end{multline}
Note that the first term in Eq. (\ref{eq:constraint}) vanishes for
$T'(0)=T'(1)$ and $T(0)=T(1)$.
By introducing
\begin{equation}
  k(\lambda, c, \mu) \equiv j^{ -1}(\lambda, c)-\mu \, n(\lambda, c)
\end{equation}
with Lagrange multiplier $\mu$ we formulate a minimization problem under
a constraint. For the optimal parameters $c,
\lambda, \mu $ which minimize the optimal inverse current
[Eq. (\ref{eq:curentmini})] and fulfill the constraint [Eq. (\ref{eq:constraintmini})],
 $\frac{\partial k}{ \partial c}$, $\frac{\partial k}{ \partial
   \lambda}$ and $n(\lambda, c)$ have to vanish. These equations cannot
 be solved analytically. Nevertheless for $c=0$ we have
 \begin{equation}
   \left.\frac{\partial k}{ \partial c}\right|_{c=0}=(\lambda-\mu)\int_0^1 \frac{dx}{\lambda T(x) \sqrt{4\lambda + \lambda^2
         T'^2(x)}}
 \end{equation}
which vanishes for $\lambda =\mu$. The partial derivative
\begin{equation}
   \left.\frac{\partial k}{ \partial \lambda}\right|_{\lambda =
       \mu, c=0}= -\frac{1}{2}\int_0^1T'(x)\:dx
\end{equation}
also vanishes for a periodic temperature profile. Thereby, we reduce the
problem to a one dimensional root search of $n(\lambda, c)|_{c=0}$ which
can easily be done numerically.

Thus, we find one solution but we cannot exclude rigorously that
there exist other solutions for this minimization problem. In our
case studies, we find (by a comparison to numerical optimization) that the 
solution with $c=0$ indeed is a global minimum. This
strongly suggests that it generally is the relevant solution.

\subsection{Perturbation theory for the divergent current}
We approximate the dependence $\lambda(\epsilon)$ for a temperature profile 
as shown in Fig. \ref{fig:templat} in the limit $\epsilon \rightarrow 0$. The constraint
[Eq. (\ref{eq:constraint})]
\begin{equation}
  \label{eq:perconstraint}
  \ln\left|1+\frac{\lambda}{8\epsilon^2}+\frac{1}{4
      \epsilon}\sqrt{\frac{\lambda^2}{4 \epsilon^2}+4
      \lambda}\right|-\sqrt{1+\frac{16
        \epsilon^2}{\lambda}}+\frac{4\epsilon-1}{\sqrt{\lambda }}+f=0
\end{equation}
relates $\lambda$ with $\epsilon$.
We define $h(\epsilon)$ by
\begin{equation}
  \label{eq:lambdae}
  \lambda(\epsilon) \equiv  4 \epsilon^2 h(\epsilon)
\end{equation}
and from Eq. (\ref{eq:perconstraint}) follows
\begin{equation}
  \label{eq:constraintf}
  \sqrt{h}\ln|1+\frac{h}{2}+\frac{1}{2}\sqrt{h^2+4h}|-\sqrt{h+4}+2 +f\sqrt{h}=\frac{1}{2\epsilon}.
\end{equation}
For $\epsilon \rightarrow 0$, the right hand side of Eq. (\ref{eq:constraintf})
diverges and thus the left hand side must also diverge, yielding
$h\rightarrow\infty$. We consider the leading terms in
Eq. (\ref{eq:constraintf})
\begin{equation}
  \label{eq:leadingf}
  \sqrt{h}(\ln |h|-1+f)=\frac{1}{2\epsilon} 
\end{equation}
and rewrite it in a self consistent equation
\begin{equation}
  h =\frac{1}{4\epsilon^2(\ln |h|-1+f)^2}.
\end{equation}
We use Eq. (\ref{eq:lambdae}) to obtain the corresponding
\begin{equation}
  \label{eq:selfcon}
  \lambda(\epsilon ) =\frac{1}{(\ln| \frac{\lambda (\epsilon )}{4 \epsilon^2}|-1+f)^2}.
\end{equation}
In perturbation theory, iteration is a standard method for algebraic
equations \cite{hinch1.1}. In this way we search for a fixed point which
is a solution for the equation. We iterate Eq. (\ref{eq:selfcon})
\begin{equation}
  \label{eq:iterationstep}
  \lambda_{n+1}(\epsilon ) =\frac{1}{(\ln|\lambda_n(\epsilon )|-2\ln|2 \epsilon|-1+f)^2}
\end{equation}
and choose 
\begin{equation}
  \label{eq:step0}
  \lambda_0=1
\end{equation}
for a good convergence. More iterations lead to
\begin{equation}
  \label{eq:step1}
  \lambda_1(\epsilon ) =\frac{1}{(2\ln|2 \epsilon|+1-f)^2}
\end{equation}
\begin{equation}
  \label{eq:step2}
  \lambda_2(\epsilon )=\frac{1}{(2\ln|2\ln|2 \epsilon|+1-f|+2\ln|2 \epsilon|+1-f)^2} 
\end{equation}
and so on.
Note that this sequence converges slowly but still gives an idea how
$\lambda(\epsilon)$ behaves. For each $\lambda$ we obtain a current from
Eq. (\ref{eq:pminicurrent}). The convergence of this sequence of currents is shown
in Fig. \ref{fig:pert}.


\end{document}